\documentclass[12pt]{article}
\usepackage{graphicx}

\newsavebox{\sboxpubnumber}
\newsavebox{\sboxpubdate}
\newcommand{\pubdate}[1]{\begin{lrbox}{\sboxpubdate}{#1}\end{lrbox}}
\newcommand{\pubnumber}[1]{\begin{lrbox}{\sboxpubnumber}{\begin{tabular}{l} #1 \\
                                 \usebox{\sboxpubdate}
                                 \end{tabular}}
                           \end{lrbox}
                           \pubblock}
\newcommand{\Title}[1]{\begin{center} {\Large #1 } \end{center}}
\newcommand{\Author}[1]{\begin{center}{ \sc #1} \end{center}}
\newcommand{\Address}[1]{\begin{center}{ \it #1} \end{center}}

\newcommand{\pubblock}{\rightline{
                        \usebox{\sboxpubnumber}}}
\newenvironment{Abstract}{\begin{quotation}  }{\end{quotation}}
\newenvironment{Presented}{\begin{quotation} \begin{center}
             PRESENTED AT\end{center}\bigskip
      \begin{center}\begin{large}}{\end{large}\end{center}
      \end{quotation}}
\newcommand{\Acknowledgements}{\bigskip  \bigskip \begin{center} \begin{large}
             \bf ACKNOWLEDGEMENTS \end{large}\end{center}}
\begin{document}

%%%%%%%%%%%%%%%%%%%%%%%%%%%%%%%%%%%%%%%%%%%%%%%%%%%%%%%%%%%%%%%%%%%%%%%%
%%
%% START EDITING HERE!
%%
%%%%%%%%%%%%%%%%%%%%%%%%%%%%%%%%%%%%%%%%%%%%%%%%%%%%%%%%%%%%%%%%%%%%%%%%
\begin{titlepage}
\pubdate{December 2001}                    %fill in the date
\pubnumber{DFTT 39/01} %preprint number(s)

\vfill
\Title{Relic neutralinos\footnote{Invited talk presented by A. Bottino}}
\vfill
\Author{A. Bottino, N. Fornengo , S. Scopel}
\Address{Dipartimento di Fisica Teorica, Universit\`a di Torino \\
         INFN, Sezione di Torino, Via Giuria 1, I-10125 Torino, Italy \\
         e--mail: bottino@to.infn.it, fornengo@to.infn.it, scopel@to.infn.it
}

\vfill
\begin{Abstract}

Phenomenology of relic neutralinos is analyzed in an effective 
supersymmetric  scheme at the electroweak scale.  It is shown 
that  current direct experiments for WIMPs, when interpreted  in 
terms of relic neutralinos, are indeed probing regions of 
supersymmetric parameter space compatible with all present 
experimental bounds. 

\end{Abstract}
\vfill
\begin{Presented}
    COSMO-01 \\
    Rovaniemi, Finland, \\
    August 29 -- September 4, 2001
\end{Presented}
\vfill
\end{titlepage}
\def\thefootnote{\fnsymbol{footnote}}
\setcounter{footnote}{0}

%%%%%%%%%%%%%%%%%%%%%%%%%%%%%%%%%%%%%%%%%%%%%%%%%%%%%%%%%%%%%%%%%%%%%%%%
% The document starts here
%%%%%%%%%%%%%%%%%%%%%%%%%%%%%%%%%%%%%%%%%%%%%%%%%%%%%%%%%%%%%%%%%%%%%%%%

\section{Introduction}

\noindent
Interest for supersymmetric relics leans on the following properties:
i) if R--parity is conserved, the Lightest Supersymmetric Particle (LSP)
is stable; ii) if colourless and uncharged, the LSP is a nice
realization of a relic
Weakly Interacting Massive Particle (WIMP) \cite{bf}. In the
present paper we assume that supersymmetry exists in Nature and that
properties (i) and (ii) hold. Thus the LSP will be our relic WIMP
candidate; actually,  after some preliminary considerations,
our attention will be mainly devoted to the case when the LSP is
provided by the neutralino.

In Sect. 2 we discuss some generic connections between WIMP relic
abundance and event rates for direct and indirect WIMP detection.
 Sect. 3 is devoted to a presentation of supersymmetric schemes. Numerical
results and conclusions are finally presented in Sect. 4.

\section{Relic abundance and detection rates}

\noindent
When we investigate a WIMP  as a relic in the Universe, the two
most crucial questions are: a) does it contribute in a significant way
to the matter content in the Universe , b) is it
detectable either directly or through indirect signals?

For a given WIMP,  its cosmological relevance and its detectability
are related, but in a way which is somewhat different from what one
might naively believe. Actually, it turns out that, in the
framework which is usually employed for the WIMP  decoupling from the
primordial plasma, WIMPs with highest detection rates might have a
relatively modest relic abundance. For a detailed discussion of these
features we refer to Ref. \cite{lathuile}; here we simply sketch the
main points.

For definiteness, our discussion is formulated in terms of relic
neutralinos, though many features apply to a larger class of WIMP
candidates, provided they share with the neutralino the property
of having an elastic scalar
cross--section off nuclei dominated by a coherent contribution
 in the range of relevance for current experiments.

Under this circumstance, the detection rate $R$ in WIMP direct
measurements  is
proportional to the product
$\rho_{\chi}  \cdot \sigma_{\rm scalar}^{(\rm nucleon)}$, where
$\rho_{\chi}$ is the neutralino local density and
$\sigma_{\rm scalar}^{(\rm nucleon)}$ is the neutralino--nucleon
cross--section.

The values to be assigned to the neutralino local density
$\rho_{\chi} = \xi \cdot \rho_l$ have to be consistent  with the values of
the neutralino relic abundance $\Omega_{\chi} h^2$.  
When  $\Omega_{\chi} h^2 \ge  (\Omega_m h^2)_{min}$, where
$(\Omega_m h^2)_{min}$ is the minimum value of $\Omega_m h^2$
compatible with halo properties, we simply set $\rho_{\chi} = \rho_l$
(i.e., $\xi = 1$).
 When  $\Omega_{\chi} h^2 <  (\Omega_m h^2)_{min}$,  the neutralino
 cannot be the  unique cold dark matter particle, thus
we assign to the neutralino a {\it rescaled} local density
$\rho_{\chi} = \rho_l \times \Omega_{\chi} h^2/(\Omega_m h^2)_{min}$
(i.e., $\xi = \Omega_{\chi} h^2/(\Omega_m h^2)_{min}$) \cite{gst}.

Taking into account these rescaling properties of the local density,
we find that $R$ behaves as follows

\begin{eqnarray}
&R& \; \; \propto \; \sigma_{\rm scalar}^{(\rm nucleon)}, \; \;  \; \; \;
{\rm when} \; \; \Omega_{\chi} h^2 \ge (\Omega_m h^2)_{min} \label{eq:1} \\
&R& \; \; \propto \; \frac{\Omega_{\chi} h^2}{(\Omega_m h^2)_{min}} \;
\sigma_{\rm scalar}^{(\rm nucleon)} \propto
\frac{\sigma_{\rm scalar}^{(\rm nucleon)}}{<
  \sigma_{ann}v >},
\;  \; \; \;
{\rm when} \; \; \Omega_{\chi} h^2 < (\Omega_m h^2)_{min}
\label{eq:2}, 
\end{eqnarray}

\noindent
 where
$\sigma_{ann}$ and $v$ are the neutralino--neutralino 
annihilation cross--section and the
relative velocity, respectively. $ <\sigma_{ann} \; v>$ denotes the
thermal average of the product $(\sigma_{ann} \cdot v)$ integrated from the freeze--out
temperature to the present--day one.
In writing Eqs. (\ref{eq:1}--\ref{eq:2}),  we used the fact
that in the standard scheme for the decoupling of a WIMP from the
primordial plasma one has
$\Omega_{\chi} h^2 \propto {<\sigma_{ann} \; v>}^{-1}$.

Now, the cross sections $\sigma_{\rm scalar}^{(\rm nucleon)}$ and
$\sigma_{ann}$, as functions of any generic coupling parameter $\zeta$, behave
similarly ({\it i.e.} they usually both decrease or increase in terms of
variations of this parameter), because of crossing symmetry.  Thus, for
instance,
$\sigma_{\rm scalar}^{(\rm nucleon)}$ and $\sigma_{ann}$ are both increasing
functions of $\tan \beta$ (see later on for definition), 
when the relevant processes are mediated by Higgs
bosons.  Usually, $\sigma_{\rm scalar}^{(\rm nucleon)}$ increases somewhat
faster than $\sigma_{ann}$, or approximately at the same rate.
Thus, due to the properties displayed in
Eqs. (\ref{eq:1}--\ref{eq:2}), the typical
behaviour of the rate $R$ may be summarized as follows: i) for small
values of $\zeta$, both $\sigma_{\rm scalar}^{(\rm nucleon)}$ and
$\sigma_{ann}$ are small (and then $\Omega_{\chi} h^2$ is
large) and $R$ grows proportionally to
$\sigma_{\rm scalar}^{(\rm nucleon)}$, ii) as the strength of the
coupling increases, in the region beyond the value
$\zeta_r$ (at which $\Omega_{\chi} h^2 = (\Omega_m h^2)_{min}$), Eq.
(\ref{eq:2}) applies: the rate $R$ still increases (though less rapidly than
before rescaling), or remains approximately flat.
This behaviour obtains up to values of the relic densities which fall
short of the cosmological interesting range by a couple of  decades.
The features just discussed for the WIMP direct detection also apply
to the indirect experiments  consisting in detecting a neutrino flux
arriving from the Sun or from the center of the Earth, due to possible
WIMP pair annihilations in these two celestial bodies. Indeed, the
size of these  neutrino fluxes depends on the product
$\rho_{\chi} \cdot \sigma_{\rm scalar}^{(\rm nucleon)}$.

On the basis of the previous arguments, we can conclude that,
in the cases of  direct
detection and indirect detection through pair annihilation in
celestial bodies,   the detectability of relic neutralinos is
  usually favoured for neutralinos of small $\Omega_{\chi} h^2$, 
{\it    i. e.} for
  neutralinos which comprise only a subdominant dark matter component.

Therefore, when
one explores the
possibility of detecting  a specific kind of WIMP, it would be
 self-defeating to restrict the analysis to the case
when the putative WIMP contributes dominantly to the matter
content in the Universe.
 Because of this fact, at variance with many
analyses by other authors, our scanning of the susy
parameter space will also include  configurations entailing
small neutralino relic abundances.
Obviously, among the different situations
which one may find in the course of the investigation, the case when
the WIMP candidate is both detectable and of cosmological interest
will represent the most rewarding occurrence.
In the last section, we show that, indeed,
  a   number of  neutralino configurations of  cosmological interest,
though disfavoured by the previous arguments,
may reach the level of detectability by current experiments.

In Ref. \cite{lathuile} it is shown that
for processes depending on pair--annihilation in the halo
the maximal rates occur for values of the relic abundance around
the value $(\Omega_m h^2)_{min}$. Subdominant neutralinos are
disfavoured for detectability by this type of signals as compared to
neutralinos with a relic abundance  around the value
$\Omega_{\chi} h^2 \simeq (\Omega_m h^2)_{min}$.

\section{Which model for supersymmetry?}

\noindent
Supersymmetry, though strongly motivated by theoretical
arguments, is not yet sustained by  experimental evidence. Only some possible
hints are available: unification of the Standard Model (SM) coupling
constants at a Grand Unification (GUT) scale \cite{chung},
Higgs events at LEP2 \cite{higgs,egno,kane,higgslep,w}, fit of precision
electroweak data improved by susy effects \cite{altarelli}. 

However, in the following we assume that supersymmetry exists in Nature and
that R-parity is conserved (thus the LSP is stable). 
 The nature of the LSP depends on the susy--breaking
mechanism and on the specific regions of the susy parameter space. We consider
here gravity--mediated schemes, and domains of the parameter space 
where the LSP is the neutralino. Extensive calculations on relic
neutralino phenomenology in gravity--mediated models 
have been performed (among the most recent references see, for instance, 
 \cite{noi,comp,probing,fcomune,acc,cn,gabr}). 

Here we refer to an  analysis which we performed in 
the Minimal
Supersymmetric extension of the Standard Model (MSSM) in a variety of
different schemes,  from those based on universal or non-universal 
supergravity, with susy parameters defined at the grand unification
scale (GUT), to an
effective supersymmetric model defined at the Electro--Weak (EW)
scale. 

\subsection{Universal and non--universal SUGRA}

  The essential elements of the MSSM are described
by a Yang--Mills Lagrangian, the superpotential, which contains all
the Yukawa interactions between the standard and supersymmetric
fields, and by the soft--breaking Lagrangian, which models the
breaking of supersymmetry.  The
Yukawa interactions are described by the parameters $h$, which
are related to the masses of the standard fermions by the usual
expressions, {\em e.g.}, $m_t = h^t v_2, m_b = h^b v_1$, where $v_i$
are the $vev$'s of the two Higgs fields, $H_1$ and $H_2$.
 Implementation of  this model within a supergravity scheme 
 leads naturally to a set of unification assumptions at a Grand
 Unification (GUT) scale, $M_{GUT}$:  
 i) Unification  of the gaugino masses:
        $M_i(M_{GUT}) \equiv m_{1/2}$,
  ii) Universality of the scalar masses with a common mass denoted by
     $m_0$: $m_i(M_{GUT}) \equiv m_0$, iii) Universality of the
     trilinear scalar couplings:
         $A^{l}(M_{GUT}) = A^{d}(M_{GUT}) = A^{u}(M_{GUT}) \equiv A_0 m_0$. 
    
    This scheme is denoted here as universal SUGRA (or simply SUGRA). The
    relevant parameters of the model at the electro--weak (EW) scale are
    obtained from their corresponding values at the $M_{GUT}$ scale by running
    these down according to the renormalization group equations (RGE). By
    requiring that the electroweak symmetry breaking is induced radiatively by
    the soft supersymmetry breaking, one finally reduces the model parameters
    to five: $m_{1/2}, m_0, A_0, \tan \beta (\equiv v_2/v_1)$ and sign $\mu$.

We remark that in this very strict scheme the phenomenology of
    relic neutralinos is very sensitive on the way in which various
    constraints (for instance those on the bottom quark mass, $m_b$,
    on the   
    top quark mass, $m_t$, and on the strong coupling
    $\alpha_s$) are implemented.

Models with unification conditions at the GUT scale
represent an  appealing scenario; however,
some of the assumptions listed above, particularly ii) and iii), are not
very solid, because, as was  already emphasized some time ago \cite{com},
universality might occur at a scale higher than $M_{GUT}\sim 10^{16}$
GeV, {\em e.g.}, at the Planck scale. 

An empirical way of taking into account the uncertainty in the
 unification scale 
 consists in allowing deviations in the
unification conditions at $M_{GUT}$.     For instance, deviations from 
universality in the scalar  masses at  $M_{GUT}$, which split 
$M_{H_1}$ from $M_{H_2}$ may be parametrized as 
$M_{H_i}^2 (M_{GUT}) = m_0^2(1 + \delta_i)$. 
This is the case of non--universal SUGRA (nuSUGRA) that we considered 
 in Refs. \cite{comp,probing,bere1}. 
Further extensions of deviations from universality in SUGRA models
 which  include squark and/or gaugino masses are discussed, for instance,
 in \cite{acc,cn}.

More recently, the possibility that
 the initial scale for the RGE running, $M_I$, might be smaller than 
 $M_{GUT}\sim 10^{16}$ has been raised,  on the basis of
 a number of string models (see, for instance, 
\cite{gabr,iban,abel} and references quoted therein). As is stressed in 
Ref.\cite{iban}, $M_I$ might be anywhere between the EW
scale and the Planck scale, with significant consequences  for the size of
the neutralino--nucleon cross section.

\subsection{Effective MSSM}

   The large uncertainties involved in the choice of the scale $M_I$ 
make the SUGRA schemes somewhat problematic: 
the originally appealing feature of a universal SUGRA with few parameters 
fails, because of the need to take into consideration the variability of $M_I$ 
or, alternatively, to add new parameters which quantify the various 
deviation effects from universality at the GUT scale. Thus, it
appears  more convenient  to work  with a phenomenological 
susy model whose  parameters are defined directly at the electroweak 
scale. We denote here this effective scheme of MSSM by effMSSM. This   
 provides, at the EW scale, a model, defined in terms of a minimum  number of 
parameters: only those necessary to shape the essentials of the theoretical 
structure of an MSSM, and of its particle content. 
Once all experimental and theoretical constraints are implemented in
this effMSSM model, one may investigate its compatibility with
specific theoretical schemes at the desired $M_I$.

In the effMSSM scheme we consider here, we impose  a set of
assumptions at the electroweak scale: 
a) all trilinear parameters are set to zero except those of the third family, 
which are unified to a common value $A$;
b) all squark  soft--mass parameters are taken  
degenerate: $m_{\tilde q_i} \equiv m_{\tilde q}$; 
c) all slepton  soft--mass parameters are taken  
degenerate: $m_{\tilde l_i} \equiv m_{\tilde l}$; 
d) the $U(1)$ and $SU(2)$ gaugino masses, $M_1$ and $M_2$, are 
assumed to be linked by the usual relation 
$M_1= (5/3) \tan^2 \theta_W M_2$ (this is the only GUT--induced
relation we are using, since gaugino mass unification appears to be
better motivated than scalar masses universality). 
As a consequence, the supersymmetric 
parameter space consists of seven independent parameters. 
We choose them to be: 
$M_2, \mu, \tan\beta, m_A, m_{\tilde q}, m_{\tilde l}, A$ and vary these 
parameters in
the following ranges: $50\;\mbox{GeV} \leq M_2 \leq  1\;\mbox{TeV},\;
50\;\mbox{GeV} \leq |\mu| \leq  1\;\mbox{\rm TeV},\;
80\;\mbox{GeV} \leq m_A \leq  1\;\mbox{TeV},\; 
100\;\mbox{GeV} \leq  m_{\tilde q}, m_{\tilde l} \leq  1\;\mbox{TeV},\;
-3 \leq A \leq +3,\; 1 \leq \tan \beta \leq 55$ ($m_A$ is the mass of
the CP-odd neutral Higgs boson).

The effMSSM scheme proves  very manageable for the susy phenomenology at the 
EW scale; as such, it has been frequently used in the literature in 
connection with relic neutralinos 
(often with the further assumption of slepton/squark mass
degeneracy: 
$m_{\tilde{q}} = m_{\tilde{l}}$). 
Notice that here we are not assuming slepton/squark mass degeneracy.

We recall that even much larger extensions of the
supersymmetric models could be envisaged: for
instance,   non--unification of the gaugino masses \cite{cn,griest},
and schemes with CP--violating phases \cite{cp}.

Here we only report results 
in  the effective scheme at EW scale, except for a few comments on
universal and non-universal SUGRA results in Sect. 4. 
For further
details we refer to Refs. \cite{lathuile,probing}.

 The neutralino is defined 
as the lowest--mass linear superposition of photino ($\tilde \gamma$),
zino ($\tilde Z$) and the two higgsino states
($\tilde H_1^{\circ}$, $\tilde H_2^{\circ}$):
$\chi \equiv a_1 \tilde \gamma + a_2 \tilde Z + a_3 \tilde H_1^{\circ}  
+ a_4 \tilde H_2^{\circ}$. 
Hereafter, the nature of the neutralino is classified in terms of a
parameter $P$, defined as $P \equiv a_1^2 + a_2^2$.  
The neutralino is called a gaugino when $P > 0.9$, a higgsino when 
$P < 0.1$, mixed otherwise.

\subsection{Constraints on supersymmetric parameters}

In our exploration of the susy parameter space, we have implemented
 the following experimental constraints:  
 accelerators data on supersymmetric
and Higgs boson searches (CERN $e^+ e^-$ collider LEP2 \cite{LEPb} 
and Collider
Detector CDF at Fermilab  \cite{cdf}); measurements of the 
$b \rightarrow s + \gamma$ decay \cite{bsgamma}. 

The new
measurement of the muon anomalous magnetic moment $a_{\mu}$
\cite{anomalous} was
recently considered by a number of authors as a sign of supersymmetry.
However, a deviation of the experimental value $a_{\mu}^{expt}$
from the SM evaluation never
exceeded a mere 1.8 $\sigma$, or even less \cite{dty}, due the uncertainties
affecting the evaluation of the hadronic contributions.
Now, in the light of the recent
clarification about the sign of the hadronic light--by--light
contribution \cite{lbl1,lbl2},
the putative  deviation effect on $a_{\mu}$ is  evaporating
away. Thus, in our analysis we do not assume that the new
determination of $a_{\mu}^{expt}$  is a sign of supersymmerty;
 we only use $a_{\mu}^{expt}$ as a constraint: the
supersymmetric  contribution to muon anomalous magnetic moment
is set to be constrained in the range:
$-200 \leq a_{\mu}^{susy} \cdot 10^{11} \leq 640$.
To establish this interval, we have combined  the results of
Refs. \cite{dh,n,j} for the hadronic vacuum polarization contributions
and
Refs. \cite{lbl1,lbl2} for the hadronic light--by--light
contributions.

As we mentioned in the Introduction, we have not restricted our 
exploration of the supersymmetric parameter
space by requiring the neutralino relic abundance $\Omega_{\chi} h^2$ 
to sit in any 
{\it a priori} chosen cosmological range.  This will enable us to
analyse our results in terms of relic neutralinos comprising 
either a dominant or a sub--dominant relic 
population. 

\section{Results and conclusions}

In what follows we present some of our results in the framework of the 
effMSSM scheme. The
evaluation of $\Omega_{\chi} h^2$ follows the procedure given in
\cite{noiom}. 

 The cross--section 
$\sigma^{\rm (nucleon)}_{\rm scalar}$ has been calculated  with the formulae
reported in Refs.\cite{noi6}. 
Important entries in  $\sigma^{\rm (nucleon)}_{\rm scalar}$ are the  
quantities $m_q <N|\bar q q |N>$, where 
the quark scalar densities $\bar q q$ are averaged over the 
nucleonic state. The values of these quantities,  derived from 
the pion--nucleon sigma term $\sigma_{\pi N}$ and other hadronic 
quantities, are affected by large uncertainties \cite{noi6}.  
The quantity $m_s <N|\bar s s|N>$, which is the most important  
term among the $m_q <N|\bar q q |N>$'s unless $\tan \beta$ is 
very small \cite{ggr}, is affected by an uncertainty factor larger than 3. 
This conclusion is reinforced by the most recent determinations of 
$\sigma_{\pi N}$ \cite{olsson,pavan}, as discussed in Ref. \cite{size}. 

The results  presented in Figs. 1--3 employ the following set 
of values for the quantities $m_q <N|\bar q q |N>$ 
(denoted 
as set 1 in Ref. \cite{noi6}): 
$ m_{l}<\bar{l}l>\; =\; 23\; {\rm MeV}, \;
  m_{s}<\bar{s}s>\; =\; 215\; {\rm MeV}, \;
  m_{h}<\bar{h}h>\; =\; 50\; {\rm MeV}$.
 For the reasons discussed above,  
all values concerning $\sigma_{\rm scalar}^{(\rm nucleon)}$ in 
Figs. 1--3 are subject to an 
increase of $\sim$ 4, when the current uncertainties in the 
$m_q <N|\bar q q |N>$'s are taken into account.

%%%%%%%%%%%%%%%%%%%%%%%%%%%%%%%%%%%%%%%%%%%%%%%%%%%%%%%%%%%%%%%%%%%%%%%%
%%
%%   use this format to include an .eps figure into your paper
%%
\begin{figure}[htb]
    \centering
    \includegraphics[height=4in]{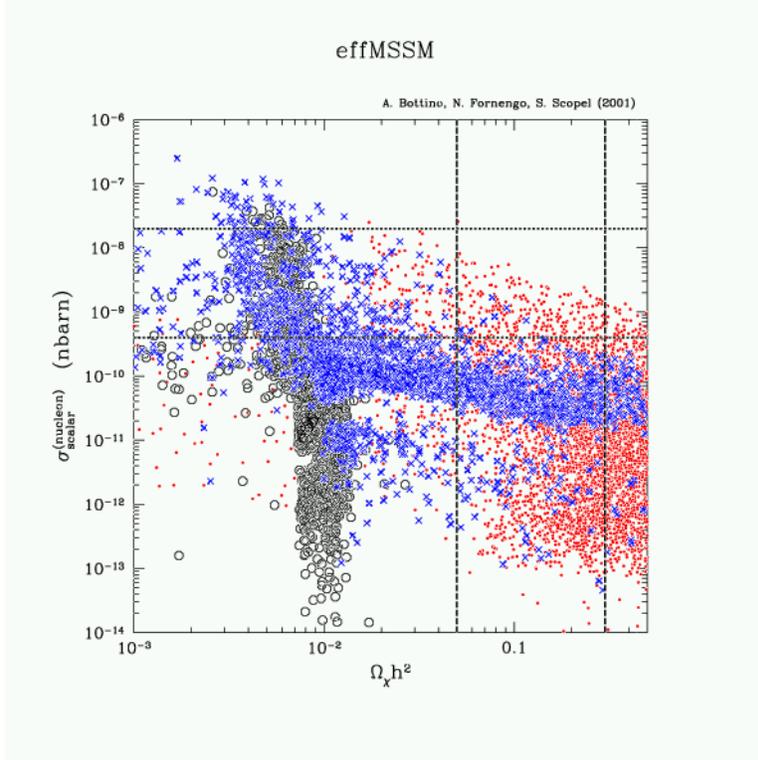}
    \caption{Scatter plot of $\sigma_{\rm scalar}^{(\rm nucleon)}$ versus 
$\Omega_{\chi} h^2$.
 $m_{\chi}$ is taken in the the interval 
$40 \; {\rm GeV} \leq m_W \leq 200 \;  {\rm GeV}$.
 The two horizontal lines bracket the sensitivity 
in current WIMP direct experiments
$4 \cdot 10^{-10} \; {\rm nbarn} \leq \
\xi \sigma^{\rm (nucleon)}_{\rm scalar} \leq
 2 \cdot 10^{-8} \; {\rm nbarn}$, for $\xi = 1$. 
 The two vertical lines denote the range 
$0.05 \leq \Omega_{m} h^2 \leq 0.3$.
Dots (crosses)
denote gaugino (mixed) configurations. 
}
%    \label{fig:cosmo}
\end{figure}
%%%%%%%%%%%%%%%%%%%%%%%%%%%%%%%%%%%%%%%%%%%%%%%%%%%%%%%%%%%%%%%%%%%%%%%%

Now let us turn to the presentation of 
our main results.  In Fig.1 
  we give the scatter plot for $\sigma_{\rm scalar}^{(\rm nucleon)}$ versus 
$\Omega_{\chi} h^2$.  The two
vertical lines denote the favorite range for $\Omega_{m}h^2$:  
$0.05 \leq \Omega_{m} h^2 \leq 0.3$. The two horizontal lines bracket
the range of sensitivity in current WIMP direct experiments 
\cite{ge,dama}, which, taking into account astrophysical uncertainties 
\cite{belli}, turns out to  be 
$4 \cdot 10^{-10} \; {\rm nbarn} \leq \
\xi \sigma^{\rm (nucleon)}_{\rm scalar} \leq
 2 \cdot 10^{-8} \; {\rm nbarn}$, 
for WIMP masses in the interval 
$40 \; {\rm GeV} \leq m_W \leq 200 \;  {\rm GeV}$. 
Fig.1 shows that the present
experimental sensitivity in WIMP direct searches allows the exploration of
supersymmetric configurations compatible with current accelerator bounds; a 
number of configurations stay inside the region of cosmological
interest. 

%%%%%%%%%%%%%%%%%%%%%%%%%%%%%%%%%%%%%%%%%%%%%%%%%%%%%%%%%%%%%%%%%%%%%%%%
%%
%%   use this format to include an .eps figure into your paper
%%
\begin{figure}[htb]
    \centering
    \includegraphics[height=4in]{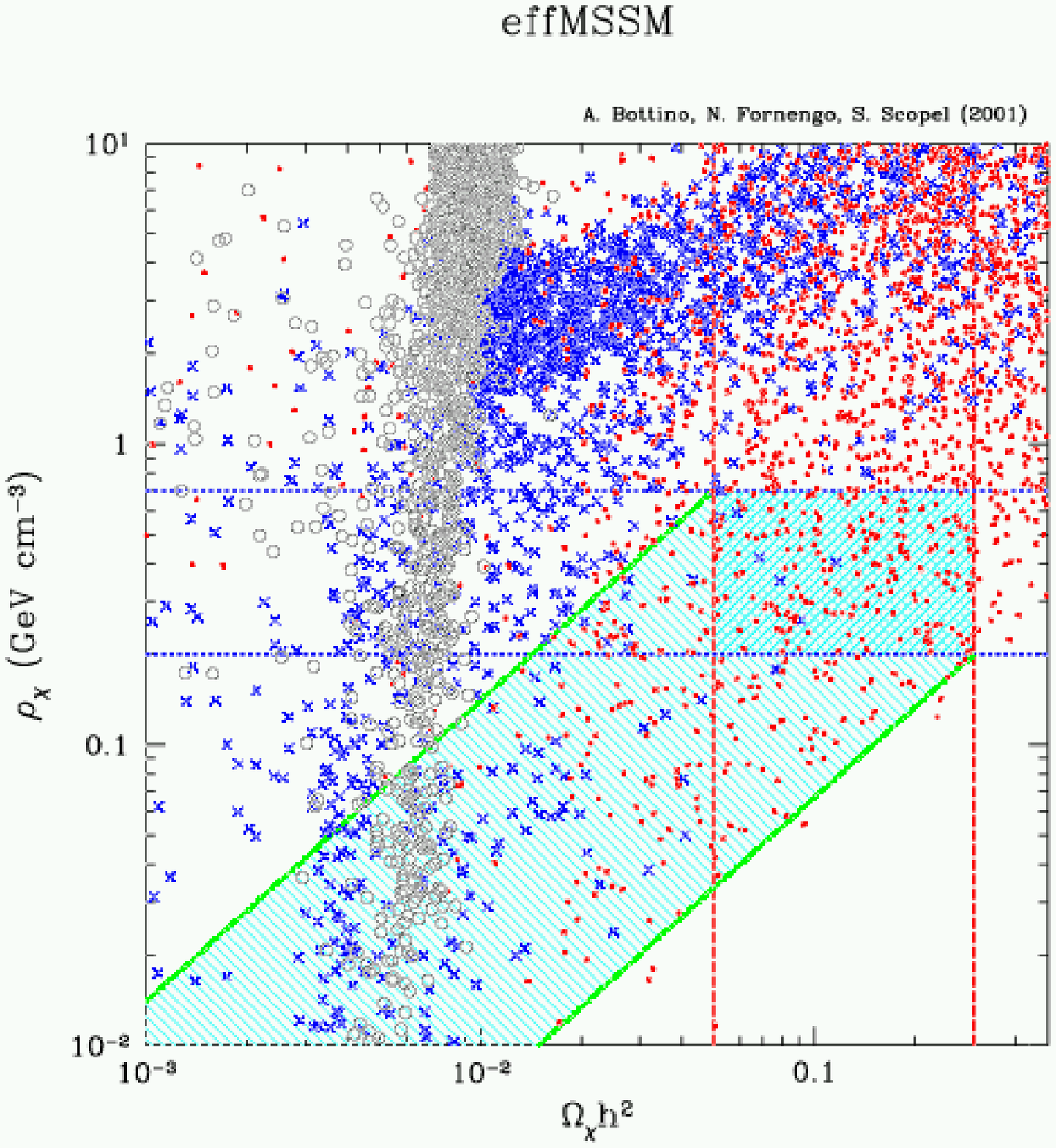}
    \caption{ Scatter plot of $\rho_{\chi}$ versus $\Omega_{\chi}h^2$.  
This plot is derived from the experimental value
  $[\rho_{\chi}$/(0.3 GeV cm$^{-3}$) $\cdot \sigma^{\rm (nucleon)}_{\rm
    scalar}]_{expt} = 1 \cdot 10^{-9}$ nbarn and by taking $m_{\chi}$ in the
  the interval 
$40 \; {\rm GeV} \leq m_W \leq 200 \;  {\rm GeV}$, according to the procedure outlined in the
  text. The two
  horizontal lines delimit the range 0.2 GeV cm$^{-3} \leq \rho_{\chi} \leq $
  0.7 GeV cm$^{-3}$; the two vertical ones delimit the range $0.05 \leq
  \Omega_{m} h^2 \leq 0.3$. The band delimited by
  the two slanted dot--dashed lines and simply hatched is the region where
  rescaling of $\rho_l$ applies.  Dots denote gauginos, circles denote
  higgsinos and crosses denote mixed configurations.
 }
%    \label{fig:cosmo}
\end{figure}
%%%%%%%%%%%%%%%%%%%%%%%%%%%%%%%%%%%%%%%%%%%%%%%%%%%%%%%%%%%%%%%%%%%%%%%%

  Once a measurement of the quantity
$\rho_{\chi} \cdot \sigma^{\rm (nucleon)}_{\rm scalar}$ is performed,
values for the local density $\rho_{\chi}$ versus the relic abundance
$\Omega_{\chi}h^2$ may be deduced by proceeding in the following way
\cite{noi6}:
1)  $\rho_{\chi}$ is evaluated as 
$[\rho_{\chi} \cdot \sigma^{\rm (nucleon)}_{\rm scalar}]_{expt}$ /
$\sigma^{\rm (nucleon)}_{\rm scalar}$, 
where $[\rho_{\chi} \cdot \sigma^{\rm (nucleon)}_{\rm scalar}]_{expt}$ 
denotes the experimental value, and 
$\sigma^{\rm (nucleon)}_{\rm  scalar}$ is calculated as indicated above;
2) to each value of  $\rho_{\chi}$ one associates the corresponding
calculated value of $\Omega_{\chi} h^2$. 
The scatter plot in Fig.2 is derived for the representative
value   
$[\rho_{\chi}$/(0.3 GeV cm$^{-3}$) $\cdot \sigma^{\rm (nucleon)}_{\rm
    scalar}]_{expt} = 1 \cdot 10^{-9}$ nbarn within the 
  annual--modulation region of Ref. \cite{dama}, and by taking 
$m_{\chi}$ in the range  
$40 \; {\rm GeV} \leq  m_W \leq 200 \;  {\rm GeV}$.

 The plot of Fig.2  shows that the most interesting region,  
{\it i.e.} the one 
with  0.2 GeV cm$^{-3} \leq \rho_{\chi} \leq $ 0.7 GeV cm$^{-3}$ 
 and $0.05 \leq \Omega_{m} h^2 \leq 0.3$ (cross-hatched region in 
the figure), is covered
   by susy configurations probed by the WIMP direct detection. 
Let us examine the various sectors of Fig.2. 
 Configurations above the upper horizontal line are
incompatible with the upper limit on the local density of dark
matter in our Galaxy and must be disregarded.
Configurations above the 
upper slanted dot--dashed line and below the upper horizontal solid line 
would imply a stronger clustering of neutralinos in our halo as 
compared to their average distribution in the Universe. This
situation may be considered unlikely, since in this case
neutralinos could fulfill the experimental range for 
$\rho_\chi$, but they would contribute only a small fraction to
the cosmological cold dark matter content.
For configurations which fall inside 
the band delimited by the slanted dot--dashed lines and simply--hatched 
in the figure,
the neutralino would provide only a fraction of the cold dark 
matter at the level of local density and of the 
average relic abundance, a situation which would be possible, for instance,
if the neutralino is not the unique cold dark matter particle
component. To neutralinos belonging to
 these configurations one 
should assign a {\it rescaled} local density. 

We remind that the scatter plot in Fig.2 refers to a representative value of
$[\rho_{\chi}$ $\cdot \sigma^{\rm (nucleon)}_{\rm scalar}]$  inside the
  current experimental sensitivity region, thus the plot in Fig.2 shows that
 current experiments of WIMP direct detection are probing relic neutralinos
  which may reach values of cosmological interest, but also neutralinos whose
  local and cosmological densities may provide only a very small fraction of
  these densities.

%%%%%%%%%%%%%%%%%%%%%%%%%%%%%%%%%%%%%%%%%%%%%%%%%%%%%%%%%%%%%%%%%%%%%%%%
%%
%%   use this format to include an .eps figure into your paper
%%
\begin{figure}[htb]
    \centering
    \includegraphics[height=4in]{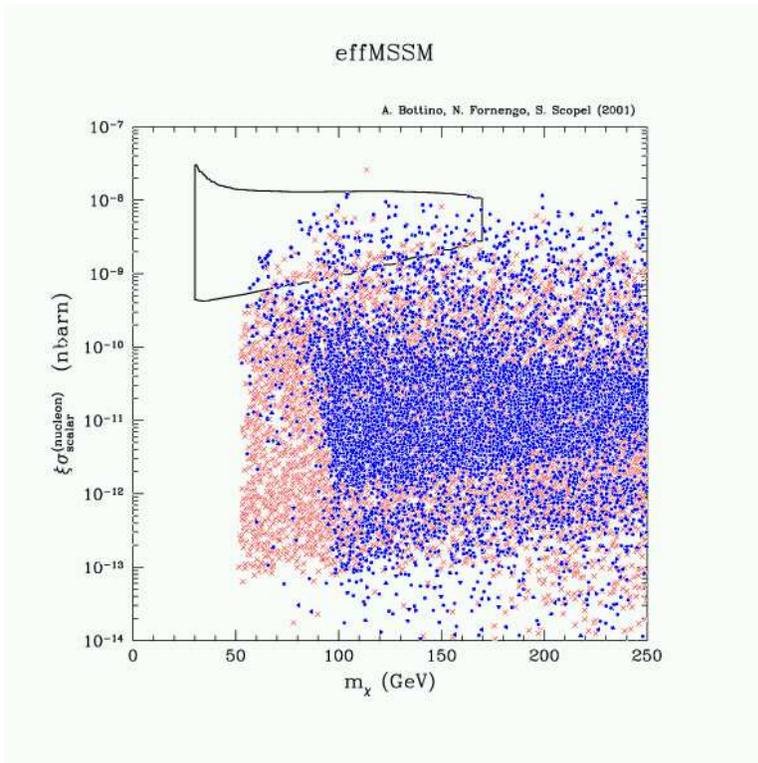}
    \caption{Scatter plot of $\xi \sigma_{\rm scalar}^{(\rm nucleon)}$ versus 
$m_{\chi}$. 
Crosses (dots) denote 
configurations with $\Omega_{\chi} h^2 > 0.05$ 
($\Omega_{\chi} h^2 < 0.05$). The solid contour denotes the 3$\sigma$ annual--modulation
region of Ref.\cite{dama} (with the specifications given in the text).
}
%    \label{fig:cosmo}
\end{figure}
%%%%%%%%%%%%%%%%%%%%%%%%%%%%%%%%%%%%%%%%%%%%%%%%%%%%%%%%%%%%%%%%%%%%%%%%

In Fig.3 we give 
 the scatter plot for the quantity $\xi \sigma^{\rm (nucleon)}_{\rm
  scalar}$ versus $m_{\chi}$. The solid line denotes the frontier of
 the 3$\sigma$ annual--modulation
region of Ref.\cite{dama}, when only the uncertainties in $\rho_l$ and in
the dispersion velocity of a Maxwell--Boltzmann distribution, but not the ones
in other astrophysical quantities, are taken into account. 
Effects due to a possible bulk rotation of the dark halo or to an
asymmetry in the WIMP velocity distribution would move this boundary towards
higher values of $m_{\chi}$ \cite{belli}.  
Our results in Fig.3 show that the susy
scatter plot reaches up the annual--modulation region of
 Ref.\cite{dama}.

In our figures only results referring to the 
effMSSM scheme are reported. For comparisons among 
results in various schemes: universal SUGRA, non--universal 
SUGRA and effMSSM, we refer to Refs. \cite{lathuile,probing}. 
As for the universal SUGRA, we only wish to remark that 
this very constrained model, combined with the 
present rather stringent experimental bounds from LEP2, 
typically entails a sizeable  suppression  of the 
neutralino--nucleon cross--section.  Whether or not this 
suppression may prevent the calculated 
$\sigma_{\rm scalar}^{(\rm  nucleon)}$ from reaching the 
region of present experimental sensitivity does depend on 
how the various constraints (typically the 
bounds on Higgs masses, on $m_t$ and $m_b$) are implemented in the 
evaluations. By way of example, it is worth mentioning that 
the explicit bounds on the quantity 
$\sin^2 (\alpha - \beta)$ ($\alpha$ being the Higgs mixing 
angle in the neutral CP--even Higgs sector) as a function of 
$m_h$ should be taken into account, rather than using a flat 
lower bound of 115 GeV for $m_h$ \cite{higgslep,w}. Neglecting 
these features in a SUGRA calculation may lead to biased 
conclusions. 

We now summarize the main points of this paper:

\begin{itemize}

\item  Most recent theoretical developments suggest supersymmetric 
schemes which notably differ from a strict model such as the 
universal SUGRA and point to the fact that this constrained
scheme should be relaxed in many instances. 
 Here, we have employed an effective MSSM 
scheme at the electroweak scale, which is particularly convenient 
to treat the relic neutralino phenomenology. 

\item  We have shown that current direct experiments for WIMPs, when 
interpreted  in 
terms of relic neutralinos, are indeed probing regions of 
supersymmetric parameter space compatible with all present 
experimental bounds. 

\item We have proved that part of the configurations probed by 
current WIMP experiments entail relic neutralinos of cosmological 
interest, and, {\it a fortiori} also neutralinos which might 
comprise only a fraction of the required amount of dark matter 
in the Universe. 

\end{itemize}

%%%%%%%%%%%%%%%%%%%%%%%%%%%%%%%%%%%%%%%%%%%%%%%%%%%%%%%%%%%%%%%%%%%%%%%%
%%
%%   use this format to include an .eps figure into your paper
%%
%\begin{figure}[htb]
%    \centering
%    \includegraphics[height=1.5in]{cosmo.eps}
%    \caption{Figure caption.}
%    \label{fig:cosmo}
%\end{figure}
%%%%%%%%%%%%%%%%%%%%%%%%%%%%%%%%%%%%%%%%%%%%%%%%%%%%%%%%%%%%%%%%%%%%%%%%

\Acknowledgements

Many of the results reported here are based on work done in
collaboration with Fiorenza Donato. 
 Financial support was partially supported by  Research
Grants of the Italian Ministero dell'Universit\`a e della Ricerca
Scientifica e Tecnologica (MURST) and of the Universit\`a di Torino 
within the {\sl Astroparticle
  Physics Project}. 
\vspace{2cm}

\end{document}